\pdfoutput=1
\documentclass{sig-alternate}

\usepackage{graphicx}
\usepackage{xspace}
\usepackage{amssymb}
\usepackage{amsmath}
\usepackage{caption}
\DeclareCaptionType{copyrightbox}
\usepackage{subcaption}
\usepackage{paralist}
\usepackage{booktabs}
\usepackage[usenames,svgnames]{xcolor}
\usepackage{pifont}
\usepackage[vlined,ruled]{algorithm2e}
\usepackage{times}
\usepackage{dsfont}
\usepackage{url}
\usepackage{multirow,bigdelim}
\usepackage{hyphenat}
\usepackage{algorithmic}
\usepackage[subtle,title=normal]{savetrees}

\newtheorem{problem}{Problem}

\newcommand{\hide}[1]{}

\newcommand{\bit}{\begin{compactitem}}
\newcommand{\eit}{\end{compactitem}}
\newcommand{\ben}{\begin{compactenum}}
\newcommand{\een}{\end{compactenum}}

\newcommand{\method}{\textsc{Flock}\xspace}

\newcommand{\ourtheory}{{\it formulation}}
\newcommand{\ouralgorithm}{{\it methodology}}
\newcommand{\ourpracticality}{{\it practicality}}

\newcommand{\baseline}{\texttt{BASELINE}\xspace}
\newcommand{\topmost}{\texttt{PRUNE-TOPMOST}\xspace}
\newcommand{\iterative}{\texttt{PRUNE-ITERATIVE}\xspace}
\newcommand{\stepwise}{\texttt{PRUNE-STEPWISE}\xspace}

\definecolor{OliveGreen}{rgb}{0,0.6,0}

\begin{document}




\title{FLOCK: Combating Astroturfing on Livestreaming Platforms}
%
%
%
%
%

\numberofauthors{1} 
%
\author{
\alignauthor
Neil Shah\titlenote{This work was done while the author was on internship at an undisclosed livestreaming corporation.}\\
       \affaddr{Carnegie Mellon University}\\
       \affaddr{5000 Forbes Ave.}\\
       \affaddr{Pittsburgh, PA, USA}\\
       \email{neilshah@cs.cmu.edu}}

\maketitle

\begin{abstract}
  Livestreaming platforms have become increasingly popular in recent years as a means of sharing and advertising creative content.  Popular content streamers who attract large viewership to their live broadcasts can earn a living by means of ad revenue, donations and channel subscriptions.  Unfortunately, this incentivized popularity has simultaneously resulted in incentive for fraudsters to provide services to \emph{astroturf}, or artificially inflate viewership metrics by providing fake ``live'' views to customers.  Our work provides a number of major contributions: (a) \ourtheory: we are the first to introduce and characterize the viewbot fraud problem in livestreaming platforms, (b) \ouralgorithm: we propose \method, a principled and unsupervised method which efficiently and effectively identifies botted broadcasts and their constituent botted views, and (c) \ourpracticality: our approach achieves over 98\% precision in identifying botted broadcasts and over 90\% precision/recall against sizable synthetically generated viewbot attacks on a real-world livestreaming workload of over \emph{16 million} views and \emph{92 thousand} broadcasts.  \method successfully operates on larger datasets in practice and is regularly used at a large, undisclosed livestreaming corporation.
\end{abstract}

%
%

%
%
%
%


\section{Introduction}

\begin{figure*}[t]
  \centering
   \begin{subfigure}[t]{0.31\textwidth}
    \centering
    \includegraphics[width=\textwidth]{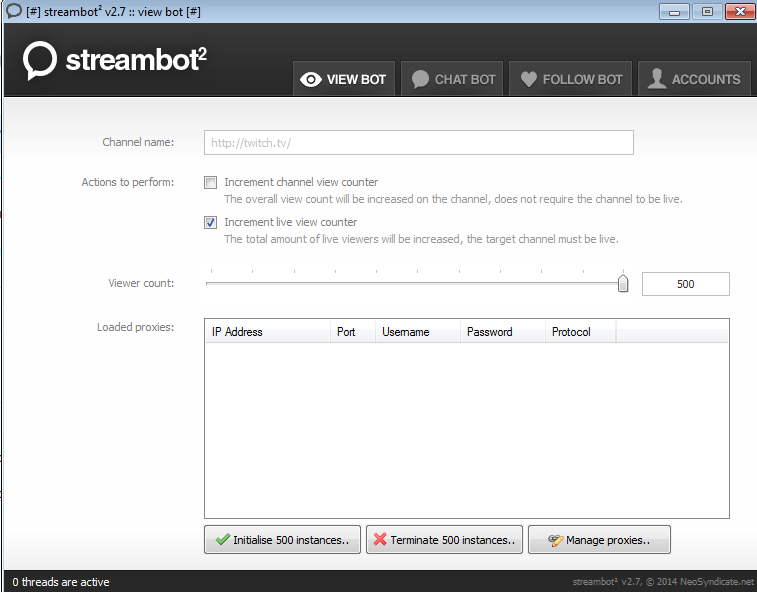}
    \caption{Viewbotting software}
    \label{fig:service_img}
   \end{subfigure}
   \quad
   \begin{subfigure}[t]{0.31\textwidth}
     \centering
     \includegraphics[width=\textwidth]{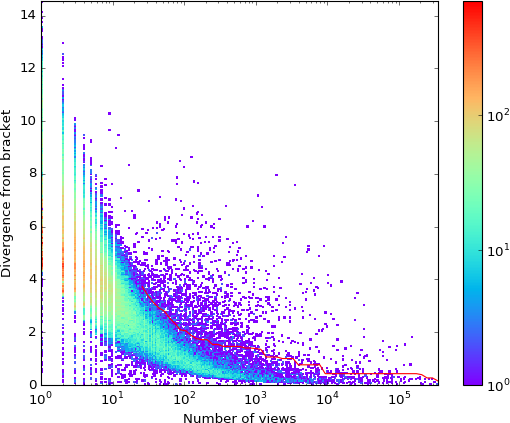}
     \caption{Viewcount-deviance plot}
     \label{fig:vc_deviance}
    \end{subfigure}
    \quad
   \begin{subfigure}[t]{0.31\textwidth}
    \centering
    \includegraphics[width=\textwidth]{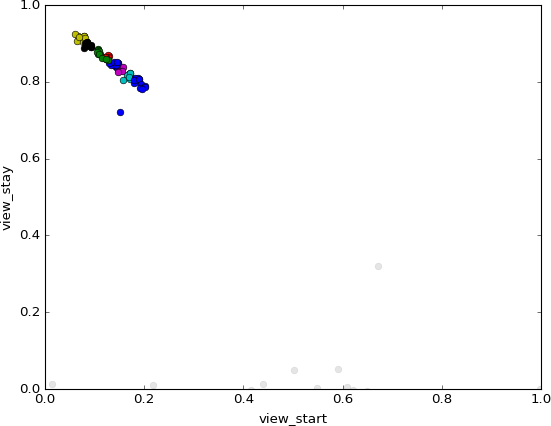}
    \caption{Broadcast-behavior plot}
    \label{fig:bcast_behavior}
   \end{subfigure}
  \caption{\textbf{\method finds botted broadcasts and their botted views.}  Figure \ref{fig:service_img} shows one of many user-controlled viewbotting tools used to astroturf broadcast viewcount from bot-provider IPs.  Figure \ref{fig:vc_deviance} shows \method's viewcount-deviance plot which plots each broadcast in the dataset by its viewcount and deviance from our model -- notice the red decision boundary separating anomalous, high-deviance broadcasts from the high density (teal) region of broadcasts.  Figure \ref{fig:bcast_behavior} shows a broadcast behavior plot for one such botted broadcast, where each point is a view's start and duration fraction through the broadcast -- notice the highly synchronized viewbot (colored) behavior where almost all views persist throughout the broadcast duration.}
  \label{fig:crown}
\end{figure*}

In recent years, livestreaming platforms have risen to provide an unprecedented level of accessible and open video content to internet users.  Livestreaming services such as Twitch, Youtube Live, and Ustream enable broadcasters to stream live video content of various types (often including electronic sports gameplay and other creative content) to an interested viewerbase who can both watch and interact with the broadcasters from their personal devices.  

Given that livestreaming has become a popular social platform for many online communities, it has simultaneously become a target for fraud by means of \emph{astroturfing}, or artificially inflating viewership and internet popularity.  As viewership is a popular target metric for recommendation and a proxy for content quality, gaming this metric offers broadcasters numerous perceived benefits including improved recommendation rankings, directory listings, monetary partnership incentives, and hopes of a resultingly larger authentic future audience.  As such, it simultaneously hinders the experience for viewers who are suggested synthetically boosted content, as well as honest broadcasters who are overlooked in favor of dishonest ones.  Numerous websites such as \texttt{viewbot.net} and \texttt{streambot.com} offer competitively priced \emph{viewbots} which can be activated and deactivated on-demand to many customers who wish to artifically inflate their ``live'' viewership. 

For these very reasons, it is important for livestreaming service providers to tackle the major issue of discerning inauthentic from authentic views.  This is exactly the problem we focus on in this work.  While the astroturfing problem in the traditional social network (followership) setting has been explored in prior literature as an unsupervised dense-subgraph mining  \cite{shah2014spotting,beutel2013copycatch,mao2014malspot} or supervised classification \cite{xiao2015detecting, freeman2013using} problem, the livestreaming (viewership) setting manifests numerous new challenges.  Firstly, the impermanence of viewbots with primarily customer-controlled (rather than operator\hyp{}controlled) parameters makes traditional dense-subgraph mining methods unsuitable -- individual customers control how many viewbots they want to use at a given time as well as when they should start and stop, so attacks will generally not involve the same bots viewing the same channels at the same times.  Secondly, the lack of sufficient ground-truth labels and indicative view-based features makes the classification task challenging.  
These differences contribute towards making the livestreaming astroturfing problem a distinct challenge from traditional settings.

In this work, to the best of our knowledge, we give the first known characterization, formulation and proposed solution to the livestream astroturfing problem.  We begin by describing prior work on both livestreaming and astroturfing frontiers.  We next describe the livest\-reaming astroturfing problem context and goals for practitioners looking to identify fraudulent behavior on livestreaming platforms both informally and formally.  We further build intuition for, and propose \method, an unsupervised, multi-step process for identifying viewbots in livestreaming settings which circumvents the aforementioned challenges.  Our \method approach works by (a) modeling broadcasts as aggregates of viewing behavior, (b) identifying anomalous (botted) broadcasts, and (c) identifying anomalous (botted) views from within these broadcasts (see Figure \ref{fig:crown}).   We next evaluate \method on a large-scale, industrial livestreaming workload and show experimental results demonstrating strong performance in practice as well as robustness against synthetic, adversarial attacks.  Lastly, we discuss means for leveraging the results from \method in practice and conclude.  Summarily, our contributions are:

\vspace{2mm}
\begin{compactenum}
 \item {\bf Problem formulation:} We describe and formalize the problem of viewbot detection in livestreaming settings.
 \item {\bf Algorithm:} We propose the \method algorithm, and unsupervised and scalable approach for finding viewbots in practice.
 \item {\bf Practical efficacy:} We demonstrate that \method achieves high precision/recall in practice and is robust and scalable.
\end{compactenum}

\section{Related Work}

As the astroturfing problem on livestreaming platforms has not previously been studied, we loosely categorize prior work into two main categories: livestreaming and astroturfing in social media.

\vspace{2mm}
\noindent {\bf Livestreaming:} \cite{veloso2002hierarchical} empirically analyzes a livestreaming workload taken from internet-accessible cameras in Brazil, and models client arrivals as a Poisson process and interest profiles as a Zipf distribution for use in a synthetic workload generation toolkit.  \cite{sripanidkulchai2004analysis} conducts a larger-scale workload analysis using data from Akamai (a major content delivery network) and additionally presents results on client diversity, lifetime and stream popularity.  \cite{wu2009queuing} studies performance of peer-to-peer (P2P) livestreaming systems using various queueing models.  \cite{liao2006anysee} presents the design of a scalable P2P livestreaming service built using inter-overlay optimization.  \cite{liu2008performance, hei2007inferring} discuss performance metrics and bounds for measuring network-quality in P2P livestreaming settings.  \cite{kaytoue2012watch} describes the prevalence of livestreaming of electronic sports (video games) and presents models to predict stream performance and popularity.  \cite{hamilton2014streaming} studies the buildup and breakdown of individual and community behaviors on Twitch streams of varying popularity.  

\vspace{2mm}
\noindent {\bf Astroturfing in social media:}   \cite{cao2012aiding, yu2006sybilguard} propose random-walk based methods which aim to leverage abnormally sparse cuts between sybil and honest regions of undirected social networks to identify sybil nodes.  \cite{prakash2010eigenspokes, shah2014spotting, jiang2014inferring, jiang2014catchsync} leverage spectral decompositions (eigendecomposition and SVD) to catch users in social networks including Twitter and Weibo who form dense subgraphs or project abnormally to low-rank subspaces. \cite{beutel2013copycatch, cao2014uncovering} detail local clustering methods on Facebook page-likes and other user actions which aim to catch synchronized behavior in the form of temporally-coherent bipartite cores in graphs.  \cite{akoglu2010oddball} finds anomalies in weighted graphs by demonstrating power-law patterns in features of graph egonets.  \cite{shah2015edgecentric, hooi2015birdnest} detail information theoretic and Bayesian approaches of identifying anomalous nodes and spammers in edge-attributed networks.  \cite{pandit2007netprobe} finds fraudulent sellers and reviewers on eBay using belief propagation when few node labels are known.  \cite{freeman2013using} uses a multinomial naive-bayes classifier on $n$-grams of account names and e-mail addresses to find spammers on LinkedIn.
\section{Background and Motivation}

Livestreaming platforms connect \emph{streamers}, or content broadcasters, with an audience of \emph{viewers}.  Each streamer broadcasts live video on their personal \emph{channel} at various times for various durations, where one continuous stream is called a \emph{broadcast}.  Broadcasts are generally associated with video games, music, television, podcasts and other creative works.  Viewers can enjoy live broadcasts by navigating to the streamer's channel during the times which he or she is \emph{live}, and subsequently watching the broadcast.  

While streaming content is just a hobby for most streamers, livest\-reaming platforms typically offer \emph{partner} status exclusively to those who consistently stream and attain a large viewership.  Partnership typically offers streamers numerous benefits including the ability to monetize from ad-revenue and paid viewer subscriptions for bonus channel content.  Streamers are thus incentivized to inflate their live viewership statistics by using viewbots to both satisfy requirements for gaining partnership status and making money, as well as to improve their rankings in livestreaming directories which viewers can browse to find popular content.  
Sufficiently successful streamers can make living wages just from streaming as a full-time job -- very popular streamers can make hundreds of thousands of dollars or more per year by means of subscriptions, donations and ad-revenue alone \cite{twitchrevenue}.  

As views must be ``live'' when streamers are broadcasting, viewbot providers offer streamers full control with regards to (a) how many viewbots they would like to use (i.e choosing 47 from 0-100), and (b) how long the viewbots should continue to watch the stream.  Upon streamer command, IPs under control of the viewbot provider emulate real viewers by sending HTTP requests for video to the livestreaming service until they are signaled to stop.  Currently, a rate limit (only allowed $k$ concurrent views per IP) is in place, meaning viewbot providers need access to $\frac{n}{k}$ IPs to successfully imitate $n$ viewers.  Given the size of the IP space from which inauthentic views can come from and transient, anonymous and streamer-controlled nature of viewbots, traditional unsupervised approaches to the astroturfing problem such as dense-subgraph detection and tensor decomposition are not suitable.  Furthermore, given that ground-truth labels are difficult to obtain and most features gleanable from HTTP requests can be spoofed, supervised models for labeling individual views require constant monitoring due to their short-lived use.

Livestreaming industry practitioners are thus faced with the difficult problem of distinguishing authentic human viewers from inauthentic viewbots.  Addressing this problem offers numerous benefits, most notably in preventing streamers who aim to cheat the system from becoming partnered, as well as limiting the inflation of perceived viewership and improving the authenticity of recommended popular content.

\section{Proposed Method: {\Large \textbf{\method}}}

We first give the intuition behind our \method approach, and subsequently describe the individual steps in greater detail.

\subsection{Intuition and Problem Formulation}

\label{sec:intuition}

As previously mentioned, labeling individual views as authentic or inauthentic from information contained in HTTP requests is difficult and adversarially error-prone.  The intuition behind \method is to take an unsupervised, offline approach which enables us to focus on viewing behaviors in \emph{aggregate} and identify behaviors that stand out from some model of \emph{normal} aggregate behavior. Specifically, we start by building a model of normal broadcast behavior, where broadcasts are considered aggregates of views.  Next, we focus on a broadcast-level analysis instead of a view-level analysis, where we examine the aggregate behavior for all viewers who watch a given broadcast and determine whether or not the overall broadcast looks suspicious  -- that is, we formulate the problem of identifying botted broadcasts as an outlier detection problem.  Broadcast-level analysis is useful, as it can account for multiple views simultaneously: while a single view which starts at 1pm and ends at 3pm on a given broadcast is not particularly suspicious, hundreds of such views are more suggestive of bot activity.  We build from this intuition and restrict our analysis to this set of botted broadcasts and try to find groups of similar views that behave like bots by starting and stopping in lockstep.  If removing such a group makes the broadcast look less suspicious in accordance with our model, then we classify those views as inauthentic.  

This hierarchical paradigm of broadcast-level and subsequent view-level analysis offers interpretability and straightforward applications to both examining streamer broadcast history for viewbotting when they apply for partnership, as well as identifying IPs commonly used for viewbot traffic.  Formally, we define our problem as follows:

\noindent \begin{problem}[Viewbot identification]
{\bf Given} the set of views $\mathcal{V}$ and corresponding functions $\alpha(v)$, $\omega(v)$, and $\chi(v)$ for all $v \in \mathcal{V}$, which represent view start time, end time, and broadcast, and the set of broadcasts $\mathcal{B}$ and corresponding functions $\alpha(b), \omega(b)$, and $\rho(b)$ for all $b \in \mathcal{B}$, which represent broadcast start time, end time, and the set of views $\left\lbrace v_1 \ldots v_k \right\rbrace \subseteq \mathcal{V}$, {\bf find} the suspicious set of broadcasts
\begin{displaymath}
\mathcal{B}_{botted} \subseteq \mathcal{B}
\end{displaymath}
and suspicious set of views 
\begin{displaymath}
\mathcal{V}_{botted} \subseteq \bigcup_{b \in \mathcal{B}_{botted}} \rho(b)
\end{displaymath}
\end{problem}

This can be broken down into several steps, each with their own subproblem, which we describe the methodology for below: (a) modeling broadcast behavior, (b) identifying botted broadcasts and (c) identifying botted views.  For clarity, please see Table \ref{tbl:symb} for reference of the recurrent symbols we use in the remainder of the section.

\begin{table}[t!]
\scriptsize
\centering
    \begin{tabular}{ll}
    \toprule
    {\bf Symbol}               & {\bf Definition}                                                                                                                         \\ \midrule
    $\mathcal{V}, \mathcal{V}_{botted} $ & set of views and botted views, resp. \\
    $\mathcal{B}, \mathcal{B}_{botted} $ & set of broadcasts and botted broadcasts, resp. \\
    $\alpha(v), \alpha(b)$ & view $v$ and broadcast $b$'s start time, resp. \\
    $\omega(v), \omega(b)$ & view $v$ and broadcast $b$'s end time, resp. \\
    $\chi(v)$ & broadcast which view $v$ watched \\
    $\rho(b)$ & views that watched broadcast $b$ \\
    $\beta(b)$ & bracket which broadcast $b$ belongs to \\
    $\zeta(t)$ & set of broadcasts in bracket $t$ \\
    $\hat{b}$ & probability distribution for broadcast $b$ \\
    $\hat{t}$ & model probability distribution for bracket $t$ \\
    $v_{start}, v_{stay}$ & start and stay feature representations for view $v$ \\
    $1_{X,Y}(v)$ & indicator noting if $v$ is in bin $(X,Y)$ \\
    $H$ & discretization parameter for broadcast duration \\
    $T$ & discretization parameter for bracket duration \\
    $K$ & IQR multiplier for broadcast decision boundary \\
    $D_{KL}(\hat{p} \parallel \hat{q})$ & KL divergence of $\hat{q}$ from $\hat{p}$ \\
    $\mathcal{I}$ & set of lockstep instances for a given broadcast $b$ \\
    \bottomrule
    \end{tabular}
\caption{\label{tbl:symb} Frequently used symbols and definitions.}
\end{table}


\begin{figure}[t!]
    \centering
    \includegraphics[width=0.5\textwidth]{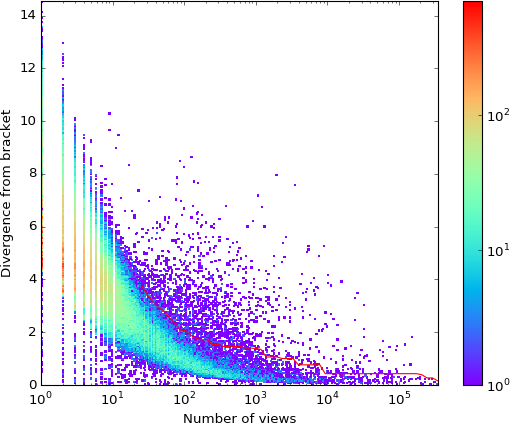}
    \caption{\label{fig:bcastdivplot} {\bf Most broadcasts deviate from brackets predictably lowly, but outliers deviate abnormally highly}.  The plot shows each broadcast as a point with viewcount on the $x$-axis and deviance between broadcast/bracket on the $y$-axis (area density denoted by color).  The red line indicates our decision boundary.}
\end{figure}

\subsection{Modeling Broadcast Behavior}

As a means towards the goal of identifying suspicious from normal broadcasts, we must first build towards a model for what normal broadcast behavior looks like.  Given the lack of ground-truth labeled data, we focus on an unsupervised model.  To build such a model, we must first identify a set of relevant features.  In this work, we avoid using descriptive (browser, country of origin, etc.) and engagement-based (chat activity, means of website navigation, etc.) features due to ease of manipulability and susceptibility to adversarial camouflage (engaging in actions that authentic viewers might do to appear human).  Instead, we focus on modeling broadcasts by the temporal features of their constituent views. 

Specifically, for each view $v$ in broadcast $\chi(v)$, we are interested in $v$'s start and end times, $\alpha(v)$ and $\omega(v)$ respectively.  Using view start time and duration is appealing, as these aspects of a view are difficult to spoof given the so-called ``mission-constraints'' of viewbot providers.  Viewbots must necessarily persist for a streamer-desired duration by their very purpose -- to boost the live concurrent viewers for an extended period of time.  

Since a view must start during, and cannot persist longer than its corresponding broadcast $\chi(v)$ respectively, it is intuitive to consider these features as fractional values rather than raw timestamps.  This is accomplished by defining the view start time and duration (``stay'' time) features ($v_{start}$ and $v_{stay}$ respectively) as

\begin{displaymath}
 v_{start} = \frac{\alpha(v) - \alpha(\chi(v))}{\omega(\chi(v)) - \alpha(\chi(v))} \, , \, v_{stay} = \frac{\omega(v) - \alpha(v)}{\omega(\chi(v)) - \alpha(\chi(v))}
\end{displaymath}

For example, if a view begins halfway through a broadcast and lasts until three-quarters of the way through, it has a $v_{start} = 0.5$ and $v_{stay} = 0.25$.  Note that this enforces the invariant $v_{start} + v_{stay} \leq 1$.

Given that each broadcast $b$ has an associated set of constituent views $\rho(b)$, we choose to model each $b$ as a random variable drawn from a joint probability distribution reflecting the frequency of empirically observed views.  To simplify representation and alleviate sparsity issues for broadcasts with small numbers of views, we discretize the otherwise continuous $v_{start}$ and $v_{stay}$ space over some number of $H$ intervals each -- $H$ is tuned empirically.  Thus, $v_{start}$ and $v_{stay}$ can each take on values from $\left\lbrace 1 \ldots H \right\rbrace$, so the post-discretization sample space has a total of $\frac{H(H+1)}{2}$ outcomes (keeping in mind the invariant) -- we will henceforth refer to $v_{start}$ and $v_{stay}$ as features defined on $\left\lbrace 1 \ldots H \right\rbrace$.

Specifically, we model each $b$ with a multinomial probability distribution $\hat{b}$ with probability masses for the outcomes defined by maximum-likelihood estimate (MLE) parameters. In this case, each view is treated as a realization of the associated random variable.  Formally, we have

\begin{displaymath}
 \hat{b}(v_{start} = X, v_{stay} = Y) = \sum\limits_{v \in \rho(b)} \frac{1_{X,Y}(v)}{|\rho(b)|}
\end{displaymath}

where $X, Y \in \left\lbrace 1 \ldots H \right\rbrace$ and $1_{X,Y}(v)$ is the indicator function which returns 1 if $v$ belongs in bin $(X,Y)$ and 0 otherwise.  While we do not formally define the indicator here in interest of space, it follows naturally as a function of $\alpha(v)$, $\omega(v)$ and $\omega(b) - \alpha(b)$.

Given this approach for modeling behavior of a single broadcast, we now aim to build a model for what normal broadcast behavior looks like.  In doing so, we build these models of normal behavior for broadcasts with different lengths separately, since viewers of different-length broadcasts are expected to behave differently.  For example, views defined with $v_{start} = 0$ and $v_{end} = 1$ (indicating a full broadcast duration view) are likely far more common for broadcasts lasting 10 minutes than they are for broadcasts lasting 10 hours.  This is due to a number of factors including time-constraints for the viewer, viewer endurance, etc.  In order to group together similar-lengthed broadcasts which we expect to be composed of similar patterns of viewer behavior, we discretize the the broadcast durations observed in our dataset into intervals spanning $T$ minutes to alleviate sparsity issues -- $T$ is tuned empirically.  We refer to these intervals as broadcast \emph{brackets}, and introduce the functions $\beta(b)$ to denote $b$'s bracket, and $\zeta(t)$ to denote the set of broadcasts in bracket $t$.  Summarily, brackets are an abstraction which serve to help us separately consider and model the behaviors of different-lengthed broadcasts.  

To model the behavior for each broadcast bracket, we take a similar approach as for individual broadcasts: we model the bracket as a joint multinomial distribution, again using MLE parameters from the empirically observed views over all constituent broadcasts in the bracket.  Formally, to compute the model distribution $\hat{t}$ for bracket $t$, we have 

\begin{displaymath}
 \hat{t}(v_{start} = X, v_{stay} = Y) = \sum\limits_{b \in \zeta(t)} \left( \sum\limits_{v \in \rho(b)} \frac{1_{XY}(v)}{\sum\limits_{b \in \zeta(t)} |\rho(b)|} \right)
\end{displaymath}

Note that the usefulness of such a model in capturing normal broadcast behavior is inherently dependent on the broadcasts in our dataset.  Our assumption is that most broadcasts are not viewbotted, and that there are far more authentic views than inauthentic views.  Over a large number of views, we expect the bracket models will be sufficiently good estimates of authentic behavior despite some viewbotted activity.  Furthermore, while individual broadcasts may have slightly different viewer behaviors due to factors such as streamer quality, directory positioning, etc., we expect that such differences will wash out over large numbers of views as well. The associated bracket distributions serve to broadly describe viewer behavior and the relative frequencies with which viewers start watching broadcasts of various lengths at different times and how long they typically watch for.  Using the bracket distributions, we can answer questions such as: ``how likely is a view which starts within seconds of a short broadcast to last for the whole duration?'' and ``how likely is the same phenomenon as broadcast duration increases?''

Now that we have a means for modeling both individual broadcasts and normal broadcast behavior (in different brackets), our next task is to discern which broadcasts are viewbotted.  

\subsection{Identifying Botted Broadcasts}

\label{sec:ibb}

We formulate the problem of differentiating authentic and viewbotted broadcasts as an outlier-detection task in which our interest is to find broadcasts that are unusually abnormal with respect to the associated bracket model.  This intuition stems from the notion that if views in a given broadcast are distributed very differently from the views in most other similar-length broadcasts, they are likely botted as they they behave in an unusual and inhuman fashion.  In order to accomplish our differentiation task, we must (a) identify a way to measure deviance between the broadcast distribution $\hat{b}$ and its associated bracket distribution $\hat{\beta(b)}$, and (b) identify a classification threshold to set for the resulting deviance scores.

There are a number of approaches for measuring statistical distance between two probability distributions, including variational distance, Hellinger distance, Kullback-Leibler (KL) divergence and others.  Of these, we choose to use the KL divergence as our distance measure as it offers nice information theoretic properties and interpretability.  The KL divergence between two distributions $\hat{p}$ and $\hat{q}$ is defined as

\begin{displaymath}
 D_{KL}(\hat{p} \parallel \hat{q}) = \sum\limits_{i} \left( \hat{p}(i) \cdot \log\frac{\hat{p}(i)}{\hat{q}(i)} \right)
\end{displaymath}

\noindent for each outcome $i$.  The divergence is defined only if $\hat{q}(i) = 0$ implies $\hat{p}(i) = 0$ in which case the summand is 0 in the limit, and is asymmetric in the sense that $D_{KL}(\hat{p} \parallel \hat{q})$ is generally not equal to $D_{KL}(\hat{q} \parallel \hat{p})$.  The KL divergence of $\hat{q}$ from $\hat{p}$ (as written above), denotes the expected number of \emph{extra} bits of information required to encode a sample from the empirically observed $\hat{p}$ distribution using a code optimized for the model $\hat{q}$ distribution rather than a code optimized for $\hat{p}$, and is thus non-negative.  It can also be interpreted as the expected log likelihood ratio between $\hat{p}$ and $\hat{q}$ when $\hat{p}$ is the actual distribution of observed data (see \cite{eguchi2006interpreting} for further detail).  Thus, the KL divergence of $\hat{q}$ from $\hat{p}$ is 0 when $\hat{p}$ and $\hat{q}$ are statistically indiscriminable, and large when $\hat{p}$ is highly unlike $\hat{q}$.  In our usecase, we consider $D_{KL}(\hat{b} \parallel \hat{\beta(b)})$ to be the appropriate measure of deviance between a broadcast and its corresponding bracket distribution.  

Given this means for measuring deviance, we next turn our attention to identifying the right threshold to set for discerning viewbotted from authentic broadcasts.  While we expect that viewbotted broadcasts will have unexpectedly high deviance from the bracket distributions, choosing this threshold is important in practical implementations to limit false positives and negatives.  In doing so, we must also keep in mind that the range of observed deviance scores will depend on the number of views for different broadcasts.  When viewcount is small, the associated distribution $\hat{b}$ will likely not be able to express the nuances of the associated bracket distribution $\hat{\beta(b)}$ very well due to sparsity issues.  Conversely, broadcasts with high viewcount should be better at approximating the associated bracket distribution because of sufficient numbers of samples.  Thus, we expect high variance in the deviance scores for low viewcount broadcasts, which should decline for higher viewcount broadcasts.  Figure \ref{fig:bcastdivplot} demonstrates this phenomenon -- the heatmap shows a single point for each broadcast in our dataset, reflecting the viewcount and the corresponding deviance (KL divergence in bits) of the bracket from the broadcast (colors indicate density as per the colorbar).  Note that most broadcasts with a certain viewcount commonly deviate from the associated bracket distributions in a predictably shrinking range of high-density, but several broadcasts have uncharacteristically high deviance as depicted in the sparse cloud of points violating the expected trend.  

These broadcasts with abnormally high deviance are exactly those which we suspect are viewbotted.  In order to automatically identify these broadcasts, we compute a moving threshold of $K$ multiples of the inter-quartile range (IQR) above the 75th percentile (3rd quartile) of samples as the outlier ``fence`` or decision boundary.  While $K = 1.5$ is typically used for normally distributed data \cite{hoaglin1986performance}, we observe that samples in each moving bin are not normally distributed -- thus, $K$ is tuned empirically in practice.  Although a $K\sigma$ boundary above the mean could also be used to similar effect, we use the 3rd quartile and IQR here as they are less sensitive to outliers.  In practice, we can also impose a threshold $U$ on the minimum viewcount we enforce a broadcast to have before we even suspect it of being viewbotted in order to avoid penalizing broadcasts with abnormally high deviance due to sparsity issues -- $U$ is tuned empirically.  These broadcasts are also far less likely to be viewbotted due to their low viewcount.  Figure 1 shows a superimposed red line indicating such an outlier fence.  

Equipped with a means of discerning viewbotted from authentic broadcasts, we next turn our attention to identifying botted views.  

\subsection{Identifying Botted Views}
\label{sec:ibv}

Broadcasts which have been viewbotted are likely composed of both authentic and botted views.  Distinguishing between these is a crucial task for identifying abusive clients as well as gaining better estimates for the amount of authentic traffic.

In order to discern between botted and authentic views, we rely on the intuition that viewbot activity results in unusually large deviance between the broadcast and bracket distributions, whereas authentic viewing behavior is more or less distributed according to (or with small deviance from) the bracket distributions.  This follows naturally from the assumption that most viewing behavior is authentic.  Thus, it stands to reason that views which result in increased deviance between the broadcast and bracket distributions are prime suspects for being botted.  As previously discussed in Section \ref{sec:intuition}, marking individual views as botted or authentic is a highly error-prone approach -- as such, labeling individual views which result in higher deviance is a risky approach.  However, again we leverage the intuition that labeling an aggregate gives higher confidence than labeling the individual -- essentially, if a similar group of views contributes to increased deviance, we can consider the constituent views in the group to be botted.  We define a ``similar'' group of views as a group whose members have temporal coherence and occur in \emph{lockstep}, or close synchrony.  A number of previous works show that lockstep behavior is a strong signal for bot activities \cite{beutel2013copycatch, cao2014uncovering, prakash2010eigenspokes}.  

Leveraging this intuition, our approach for identifying botted views is to first identify instances of lockstep behavior (groups of similar views in a given broadcast), and mark these instances as botted if pruning their constituent views from the broadcast results in a smaller deviance between the pruned/altered broadcast and the associated bracket distribution.  Formally, we aim to solve the following optimization problem for each outlier broadcast $b$:

\begin{displaymath}
 \min_{b' \in 2^{\mathcal{I}}} D_{KL}(\hat{b'} \parallel \hat{\beta(b)})
\end{displaymath}

\noindent where each of the $|\mathcal{I}|$ instances corresponds to an exclusive subset of views $\rho(b)$ such that the intersection between any two instances is the emptyset, $2^{\mathcal{I}}$ denotes the powerset of $\mathcal{I}$ (such that each element corresponds to a subset of instances), and $b'$ and $\hat{b'}$ are the pruned broadcast and resulting empirical probability distribution formed by the views in the corresponding subset of instances.  Intuitively, $\mathcal{I}$ gives a hard partitioning of $\rho(b)$ into $|\mathcal{I}|$ instances, and we are after the most seemingly authentic combination of instances, where the objective function for authenticity is the KL divergence between the pruned broadcast and associated bracket distribution.  This problem has multiple associated challenges.  Firstly, we must identify a means of partitioning the set of views $\rho(b)$ into some number of $|\mathcal{I}|$ instances. Secondly, as finding the exact solution once $\mathcal{I}$ is fixed is still a combinatorial minimization task, we must find an efficient and scalable means for tackling the problem.

For the first task, we can construe instances as clusters of views in the 2D space spanned by $v_{start}$ and $v_{stay}$.  Clustering of similar data is a common data mining task, and we can select from a number of available clustering algorithms including hierarchical clustering, DBSCAN, K-means and more (see \cite{berkhin2006survey} for an overview).  Of these, K-means is commonly chosen due to its scalability and well-understood squared loss criterion.  However, one notable caveat of K-means is the requirement of a user-specified number of clusters.  Knowing this parameter a priori is difficult in our usecase and many others.  Thus, we instead use an algorithm similar to X-means \cite{pelleg2000x}, which automatically infers a suitable number of clusters by aiming to minimize the Bayesian Information Criterion (BIC), a function of log-likelihood penalized by model complexity.  The approach starts with all data points in a single cluster, and iteratively splits clusters into smaller clusters if the split reduces the BIC.  While the original X-means uses traditional K-means in the inner-loop, we instead use a much faster minibatch variant proposed in \cite{sculley2010web} which is shown to give only marginally worse solutions than traditional K-means.  This approach outputs a hard-partitioning of the views into a suitable number of instances.  While the clustering approach is heuristic, an intuitively good clustering can generally be achieved over a few initializations.

The next challenge is to actually choose the subset of $\mathcal{I}$ which minimizes the above objective.  As the task is to find the subset which has a minimum divergence over all the subsets, the problem is clearly combinatorial and would require checking $2^{|\mathcal{I}|}$ subsets for a broadcast partitioned into $|\mathcal{I}|$ instances.  As this becomes computationally expensive quickly even for small values of $|\mathcal{I}|$ and given that our objective is non-submodular, we resort to heuristics to approximate the solution.  Below, we propose several greedy heuristics for pruning the botted instances.
\vspace{2mm}
\begin{compactitem}
 \item {\bf \topmost} -- Rank each instance $\mathcal{I}_i$ in decreasing order according to the resulting reduction in deviance if it is pruned from the original broadcast $b$.  Prune out the single instance which results in maximal reduction.  If no such instance exists, stop.
  
 \item {\bf \iterative} -- Rank each instance $\mathcal{I}_i$ in decreasing order according to the resulting reduction in deviance if it is pruned from the original broadcast $b$.  Attempt to prune each of the instances in that order, removing those which result in reduction from the current best.  A full pass over instances is considered 1 iteration.  Repeat this iteration until convergence, starting with the current best broadcast from the previous iteration.  If no instance is removed between the start of successive iterations, stop.
 
 \item {\bf \stepwise} -- Rank each instance $\mathcal{I}_i$ in decreasing order according to the resulting reduction in deviance if it is pruned from the original broadcast $b$.  Prune out the single instance which results in maximal reduction.  Repeat this step until convergence, starting with the current best broadcast from the previous step.  If no instance is removed between the start of successive steps, stop.
 
\end{compactitem}

\vspace{2mm}

Though the heuristics each offer different means of pruning botted views, they all start by considering the full set of instances $\mathcal{I}$ as the initial ``best'' solution, and greedily removing individual instances which result in an improvement in the objective function.  Specifically, at different points in each heuristic, the test $D_{KL}(\hat{b}_{current} \parallel \hat{\beta(b)}) - D_{KL}(\hat{b}_{proposed} \parallel \hat{\beta(b)})$ is conducted in the scenario that some instance is removed from the current best broadcast $b_{current}$ resulting in the candidate broadcast $b_{proposed}$, and if the result is $\geq 0$, the instance is removed from $b_{current}$ and the instances forming $b_{proposed}$ becomes the new best solution.

\topmost involves ranking the instances only once, and pruning the single instance which improves the objective the most.  This objective is motivated by the notion that most bot behavior might be naive and viewbot attacks may typically only span one instance per broadcast.  \iterative and \stepwise both involve pruning multiple instances if doing so improves the objective, but vary in the extent of greediness.  \iterative ranks the set of remaining instances at the start of each iteration based on the current best broadcast at the start of that iteration.  However, in reality, as soon as an instance is pruned during an iteration, the true ranking for the next best candidate instances to prune may change.  It is furthermore possible that the removal of an instance in one iteration may induce or preclude the removal of another instance in a future iteration.  The \iterative heuristic does not dynamically update the ranking upon each successful instance pruning but rather on an iteration-level, based on the notion that this iteration-level ranking may be sufficient in practice.  \stepwise conversely \emph{does} dynamically update the ranking after each step, or successful pruning of an instance.  This way, the best candidate instance for pruning is chosen at each step at the cost of more frequent ranking computations compared to \iterative if multiple instances are to be pruned.  These heuristics have various expected tradeoffs in computational efficiency and objective minimization efficacy.  We examine their practical performance in Section \ref{sec:eval}.  

In conjunction with previous tasks, this gives us a means of identifying both viewbotted broadcasts and their botted views.  See Algorithm \ref{alg:flock} for pseudocode of our proposed \method approach.

\begin{algorithm}[t!]
\scriptsize
\caption{\label{alg:flock} \method}
\begin{algorithmic}[1]
\STATE {\bf Model Broadcast Behavior}: Aggregate per-broadcast and per-bracket views and transform into $v_{start}, v_{stay}$ space.  Build viewership models $\hat{b}$ and $\hat{t}$ for each broadcast $b$ and bracket $t$ according to multinomial MLE parameters.
\vspace{0.5mm}
\STATE {\bf Identify Botted Broadcasts}: Compute deviance of each broadcast's bracket distribution $\hat{\beta(b)}$ from the broadcast distribution $\hat{b}$, and generate viewcount-deviance plot as in Figure \ref{fig:bcastdivplot}.  Generate decision boundary by binning viewcount logarithmically and computing a $Q_3 + K \cdot IQR$ fence per bin.  Mark broadcasts with excessively high deviance as botted.  
\vspace{0.5mm}
\STATE {\bf Identify Botted Views}: For each broadcast $b$, partition $b$ into a set of instances/clusters $\mathcal{I}$.  Attempt to solve the minimization problem $D_{KL}(\hat{b'} \parallel \hat{\beta(b)})$ where $b'$ is chosen from the powerset of instances $2^{\mathcal{I}}$.  Use the most suitable heuristic from \topmost, \iterative and \stepwise.
\end{algorithmic}
\end{algorithm}

\section{Experimental Results}

\label{sec:eval}

In this section, we empirically evaluate our proposed \method approach's effectiveness in detecting viewbots.  Firstly, we briefly describe the dataset used in the evaluation.  Next, we describe several experiments in order to (a) measure success in identifying viewbotted broadcasts, (b) evaluate comparative performance of our view pruning heuristics, (c) demonstrate effectiveness in reliably discerning authentic from botted views, and (d) consider implications in the presence of a well-informed adversary.  These are detailed in the below subsections.

\subsection{Data Description}

For the experiments below, we use data from a large, undisclosed livestreaming corporation spanning an 8 hour timeframe from early May 2016.  Our dataset consists of 92,044 broadcasts and an associated 16,280,308 views.  For each broadcast, we have information about the broadcasting channel and the start and end times of the broadcast.  For each view, we have information about the client IP, start and end times and the destination channel.  From these, we can associate views with broadcasts.    

\subsection{Broadcast Classification}

\begin{figure*}[t!]
    \centering
    \begin{subfigure}{0.49\textwidth}
        \centering
        \includegraphics[width=\textwidth]{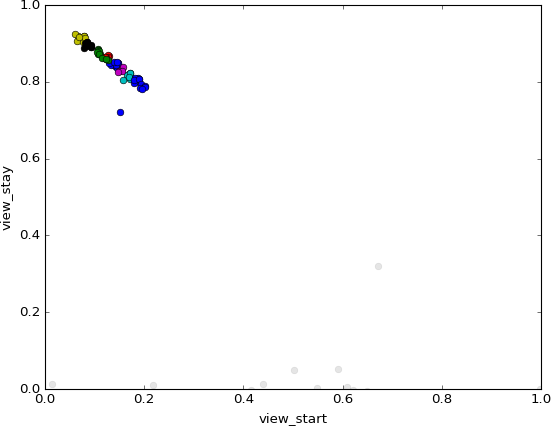}
        \caption{\label{fig:bottedbcasts_a} 76/89 broadcast views began soon after the start and persisted for the full duration.  Note that the other 13 views are much sparser and shorter.}
        \end{subfigure}
    \hfill
    \begin{subfigure}{0.49\textwidth}
        \centering
        \includegraphics[width=\textwidth]{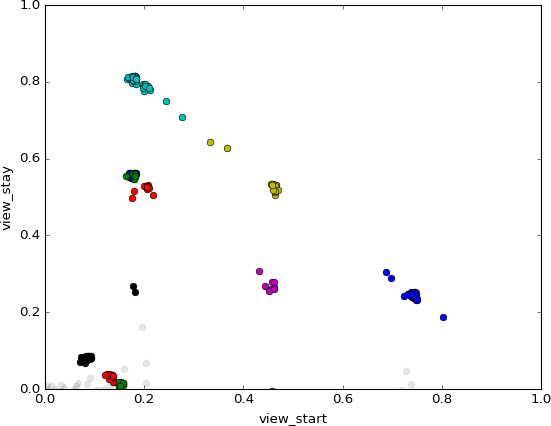}
        \caption{\label{fig:bottedbcasts_b} 201/239 broadcast views are manifested in dense clusters indicating bots in lockstep.  The teal/blue instances on the diagonal contain 50 views each.}
    \end{subfigure}
    \caption{\label{fig:bottedbcasts} {\bf Viewbots creates dense clusters of activity in the $(\mathbf{v_{start}}$, $\mathbf{v_{stay}})$ space.}  Two fraudulent broadcast behavior plots are shown with suspected botted instances and their constituent views plotted in opaque color and unsuspected authentic views in translucent grey.}
\end{figure*}

The first important consideration for evaluating effectiveness is determining whether our intuition that outlier broadcasts (the sparse cloud in Figure \ref{fig:bcastdivplot}) are in fact viewbotted is correct.  Unfortunately, calculating precision is difficult given the lack of ground truth.  In lieu of existing labels, we resort to manual labeling.  We randomly sample 100 broadcasts each from the outlier and non-outlier regions according to the decision boundary, intermix them, and manually label each broadcast using criteria based on (a) IP/ASN entropy (whether many temporally similar views come from a limited subset of IP prefixes or ASNs), and (b) whether there are high-density clusters of seemingly out-of-place views.  

Several such examples of suspected viewbotted broadcasts with the associated suspected views are depicted in Figure \ref{fig:bottedbcasts}.  In both cases, large fractions of broadcast views are concentrated in tightly-knit clusters which upon further inspection were from a small number of shared IP prefixes under a few lesser known international ASNs.  Figure \ref{fig:bottedbcasts_a} shows a rather simplistic type of viewbotting, in which bots join incrementally and quickly soon after a broadcast begins, and persist until the end of the broadcast.  Conversely, Figure \ref{fig:bottedbcasts_b} shows a more complex type of viewbotting in which the streamer is tuning the bot viewership over the duration of the broadcast as he or she desires.  Intuition suggests this might occur in three cases: (a) the streamer is simply experimenting with a viewbot provider's tool, (b) the streamer feels they have activated too many bots or too few bots initially, but massages the number over time in order to try and appear legitimate, (c) malfunctions are occuring on the viewbot provider's side, resulting in unusual termination.  Retroactive analysis of the live concurrent viewers counter for this broadcast indicates sharp/sudden spikes and drops (typically of 20 viewers at a time) as expected given the sudden starting/stopping of various numbers of bots indicated in the plot.  

The results of our labeling experiment indicate 98\% positive and 99\% negative precision.  That is, 98\% of outlier broadcasts and 99\% of non-outlier broadcasts were labeled viewbotted and non-viewbotted respectively.  Recall numbers can unfortunately not be calculated due to the unbounded number of false negatives.  These figures suggest that our proposed decision boundary and unsupservised classification approach is able to make a highly accurate distinction between viewbotted and non-viewbotted broadcasts.  As in most threshold-based classification settings, the $K$ (IQR multiplier used for computation of the decision boundary) can be increased to further limit false positives while catching fewer true positives, or decreased with an inverted effect if desired.

\subsection{View Pruning Heuristics}

\label{sec:eval_vph}

\begin{table}[t]
\scriptsize
\setlength{\tabcolsep}{2pt}
\centering
\begin{tabular}{@{}llllll@{}}
\toprule
 & \textbf{\begin{tabular}[c]{@{}l@{}}Original $\mathbf{D_{KL}}$\\ (avg. bits)\end{tabular}} & \textbf{\begin{tabular}[c]{@{}l@{}}Pruned $\mathbf{D_{KL}}$\\ (avg. bits)\end{tabular}} & \textbf{\begin{tabular}[c]{@{}l@{}}MAD \\ (bits)\end{tabular}} & \textbf{\begin{tabular}[c]{@{}l@{}}MAPD\\ (\% bits)\end{tabular}} & \textbf{\begin{tabular}[c]{@{}l@{}}Runtime\\ (sec)\end{tabular}} \\ \midrule
\textbf{\baseline} & 3.28 & 3.28 & 0 & 0 & 0 \\
\textbf{\topmost} & 3.28 & 2.87 & 0.408 & 12.26 & 1277.65 \\
\textbf{\iterative} & 3.28 & 2.25 & 1.035 & 31.27 & 2655.51 \\
\textbf{\stepwise} & 3.28 & 2.23 & 1.049 & 31.86 & 5119.04 \\ \bottomrule
\end{tabular}
\caption{\label{tbl:viewpruning} Quantitative evaluation of view pruning heuristics versus baseline (no pruning).}
\end{table}

As previously discussed, the per-broadcast view pruning minimization task proposed in Section \ref{sec:ibb} is combinatorial, so we propose the use of several greedy heuristics with varying tradeoffs between presumed optimization performance and scalability.  In this section, we evaluate the heuristics in these terms.  \topmost costs $\mathcal{O}(H^2 \cdot (|\mathcal{I}| + 1))$ for ranking the $|\mathcal{I}|$ instances and computing the KL divergence each time, and pruning a single instance.  \iterative costs $\mathcal{O}(H^2 \cdot 2|\mathcal{I}| \cdot p)$ for $p$ pruning iterations where each iteration involves ranking and trying to prune each of the $|\mathcal{I}|$ instances (we find $p$ is generally very small in practice, $\leq 5$).  \stepwise costs $\mathcal{O}(H^2 \cdot (\mathcal{I} + 1) \cdot s)$ for $s$ pruning steps, each of which involves running \topmost.  Generally, \stepwise is slower that \iterative, which is slower than \topmost.  However, the heuristics have inverse performance on the minimization task.  

Table \ref{tbl:viewpruning} gives empirical results which demonstrate these findings in comparison to a null baseline, in which no pruning is performed.  For each heuristic, we prune each outlier broadcast in our dataset and measure the initial average deviance ($D_{KL}$, in bits) from the broadcasts and their associated bracket distributions.  We next prune each of the broadcasts using the various heuristics and measure the post-pruning average deviance per broadcast, mean absolute (percentage) deviation (MAD and MAPD) and total runtime over all broadcasts (over 5 initializations).  Note that here, higher is better for MAD and MAPD, indicating a larger drop in the deviance $D_{KL}$.  Note that given $D_{KL}$'s interpretation as a log likelihood ratio, an average improvement of even 1 bit corresponds to doubling agreement between distributions.  While minimization performance is inversely related to the computational performance of the heuristics, the benefit of \iterative over \topmost is quite large (200\%) compared to \stepwise over iterative (3\%).  Given the much worse scaling of \stepwise in situations where many instances are pruned, we choose to use \iterative in practical implementation as it minimizes the objective almost equally well.

\subsection{View Classification}

\begin{figure}[t!]
    \centering
    \includegraphics[width=.5\textwidth]{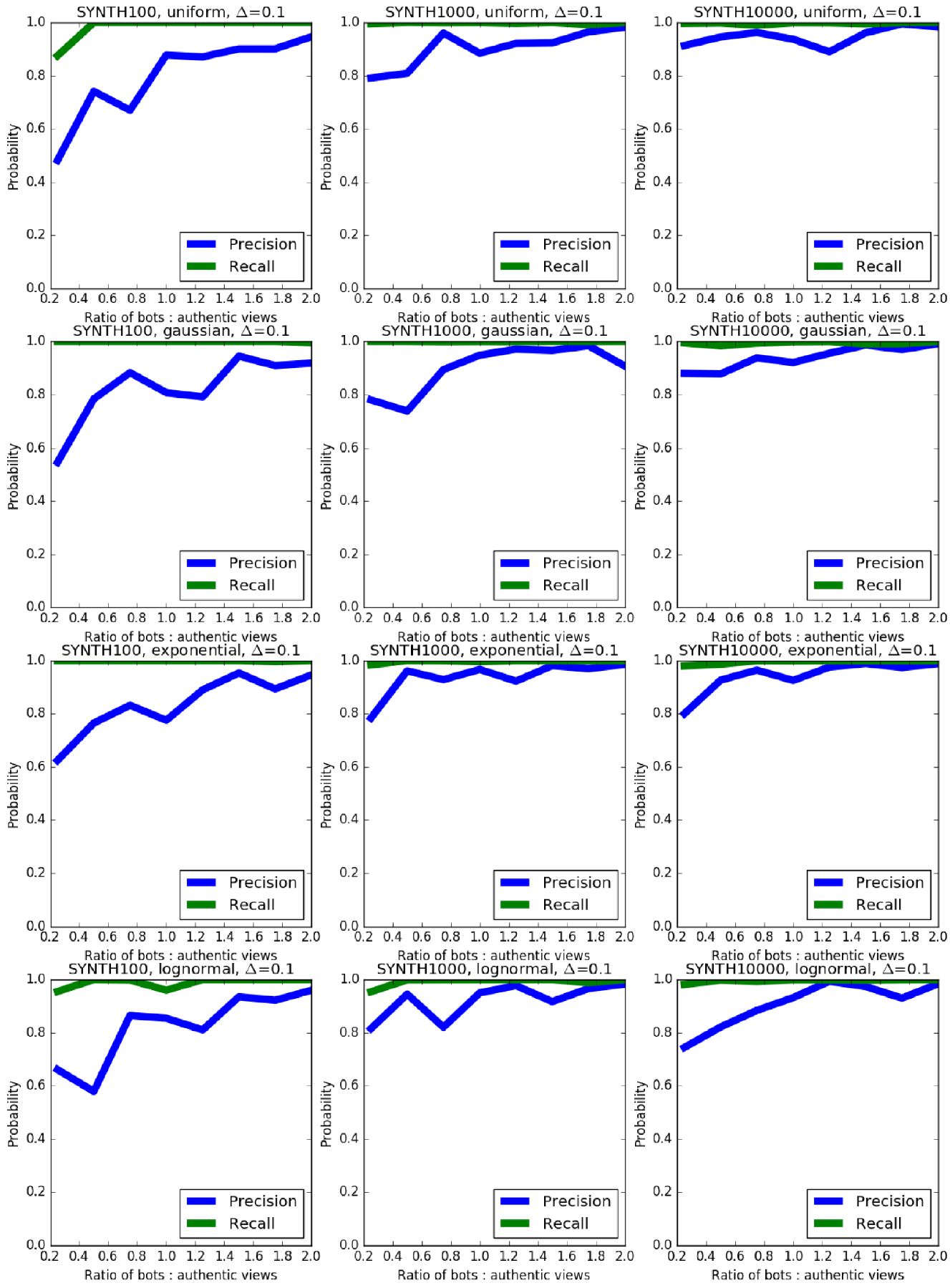}
    \caption{\label{fig:synth_attacks} \textbf{\method achieves high precision/recall in discerning synthetically botted views from authentic views.}}
\end{figure}

As getting ground truth labels on a per-view basis is even more difficult to obtain than on a broadcast-level, we turn to synthetic experiments to evaluate \method's performance in discerning botted views from authentic ones.  To do so, we simulate botted broadcasts by varying numbers of authentic views and botted views by sampling the two types from various distributions.  Specifically, we sample authentic views from a real bracket distribution (multinomial) and jitter them with Gaussian noise.  Subsequently, for each broadcast, we sample the first botted view's $v_{start}$ and $v_{stay}$ time uniformly over the feasible space, and subsequent botted view interarrival times (IAT) and intertermination times (ITT) from a number of distributions which viewbot providers might use to engineer real attacks.  As Figure \ref{fig:bottedbcasts} indicates that botted views are almost in lockstep and delivered over a short amount of time, we artifically deliver and terminate these botted views in accordance over 10\% of broadcast duration respectively -- we refer to this attack scale parameter as $\Delta$.  Intuitively, $\Delta = 0.1$ indicates that botted views were delivered over a 10\% timeframe, and then all terminated in a different 10\% timeframe. This is a realistic setting from our observations.  We found results were not sensitive to $\Delta$.  

Figure \ref{fig:synth_attacks} summarizes precision and recall results from experiments on 96 combinations of the following 3 parameters: authentic viewcount (chosen from 100, 1000 and 10000), proportion of botted views (chosen from $0.25, 0.5 \ldots 2.0$ -- functions as a multiplier on the number of authentic views), and viewbot attack distribution (IAT and ITT chosen from uniform, Gaussian, exponential and lognormal distributions). Precision and recall figures were averaged over 5 runs of each experiment.  The results show consistently high ($\geq 0.95$) recall in almost all cases, and high ($\geq 0.9$) precision for broadcasts with especially high levels of botted activity.  The precision tends to increase with higher numbers of botted views as the clustering process is better able to separate the botted views from authentic ones and successfully prune them.  Precision also tends to be better for broadcasts with higher number of authentic views as the relative fraction of authentic views captured in an instance of mostly botted views (false positive rate) is lower.  These results are especially promising for practical scenarios as in Figure \ref{fig:bottedbcasts} in which we observe that a large fraction of total views are botted.  

\subsection{Adversarial Implications}

In previous experiments, we operated from the assumption that botted views are delivered in lockstep based on practical observations.  But, what are the implications of our approach in a setting with an adversary who has complete knowledge of the bracket distributions he must emulate to appear normal?  Note that this is a very strict setting -- in order to have the correct bracket distribution to sample from, the adversary must a priori know (a) broadcast duration (b) internal records on viewerbase behavior.  As such, we do not expect this level of sophistication from an attacker, but it is a useful thought experiment.

We conduct an experiment in which the adversary acts to minimize risk of being caught by sampling directly from the target bracket distribution (provably optimal, as deviance between broadcast and bracket is 0 in the limit).  Since adversaries are IP constrained, we aim to evaluate the IP cost implications upon adopting this strategy.  Specifically, we calculate how many more IPs are required to reach the same average concurrent viewer count for \method with a rate limit compared to a standalone rate limit.  We estimate this figure by computing the expected value of a random variable with outcomes corresponding to the relative overhead in maximum concurrent viewers required under both regimes per bracket and probability corresponding to the empirical bracket frequency.  Our results indicate that using \method with a rate limit incurs a $40\%$ overhead in IP addresses required to viewbot compared to naive rate limiting -- intuitively, viewbot providers need 40\% more IPs to add ``noise'' resembling real views rather than simply activating bots in lockstep like in Figure \ref{fig:bottedbcasts}.  Thus, we conclude that even against a knowledgeable adversary, our approach significantly increases the cost of viewbotting.

\subsection{Scalability}

\begin{figure}[t!]
    \centering
    \includegraphics[width=0.4\textwidth]{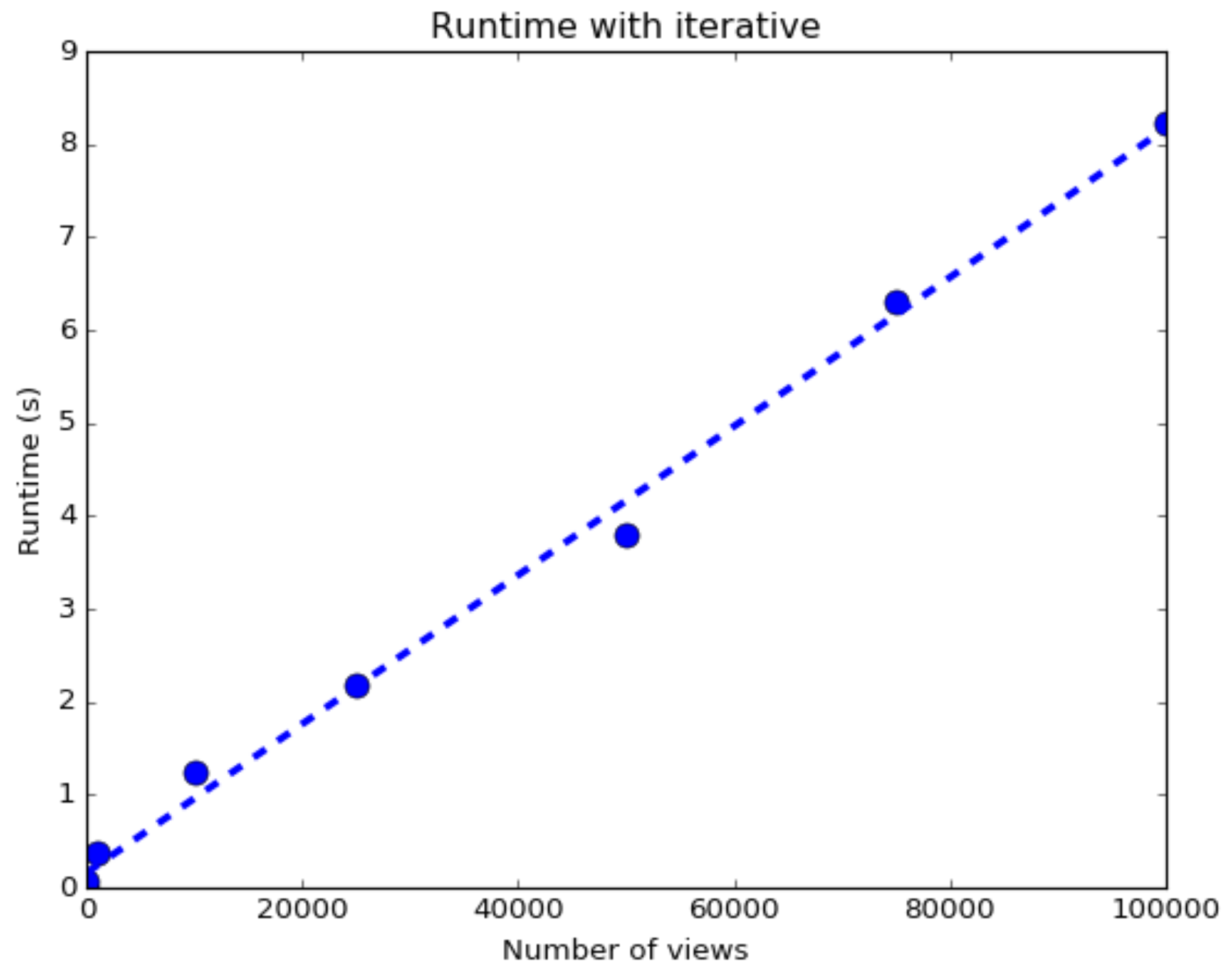}
    \caption{\label{fig:scaling} \textbf{\method scales linearly in the number of views.}}
\end{figure}

\method's analysis phase begins by clustering each of the outlier broadcasts in $\mathcal{B}_{botted}$.  As X-means has been shown to scale better than traditional K-means, we consider the time complexity for clustering the views of each broadcast $b$ to be at worst $\mathcal{O}(|\rho(b)| \cdot |\mathcal{I}| \cdot d \cdot c)$, for $|\rho(b)|$ views, $|\mathcal{I}|$ instances, $d$ dimensions (2 in our case) and $c$ clustering iterations.  Upon clustering, we use \iterative to prune suspected botted instances and views from each broadcast -- see Section \ref{sec:eval_vph} for complexity analysis of the heuristics.  Figure \ref{fig:scaling} shows the linear scaling of \method on a number of synthetically generated broadcasts with varying viewcount.

\section{Discussion}

\method is an unsupervised, offline (post hoc) approach for detecting botted broadcasts and views in livestreaming.  In practice, the results can be leveraged in a number of ways.  Firstly, \method can be used daily on newly collected view and broadcast data.  
The results of these daily runs can provide history of a streamer's broadcasts as well as history of each seen IP's previous views along with \method's associated ruling of (in)authenticity.  The former can be used to aid in making parternship decisions by filtering fraudulent applicants with a history of botted broadcasts.  The latter can be used to better estimate true past viewership, and adjust future live viewcounts using IP-based penalties reflecting the prevalence of previously botted activity from the IP.  It can also be used to identify viewbot browser, service provider and environment signatures decipherable from the received HTTP requests and proxy ground ``truth'' for a supervised scheme.  

While not the focus of this work, \method could be used in other contexts such as detecting fake ad clicks and video views where ground truth is difficult to obtain.  While the feature space may be different, a similar approach of finding lockstep instances which violate ``normality'' assumptions appears promising.
\section{Conclusion}

Online livestreaming has become a prevalent means for individuals to broadcast creative content to the masses.  Streamers have high incentive to gain viewership in order to promote their brand, gain status and earn a living through ad revenue, donations and subscriptions.  In this work, we are the first to focus on the problem of astroturfed, or artificially inflated viewership on livestreaming platforms.  We begin by characterizing the livestreaming context and formalizing the problem setting for identifying fake views.  To this end, we propose \method, a principled and scalable method for identifying botted broadcasts and views in an unsupervised fashion.  \method works by first discerning broadcasts with highly abnormal and seemingly inhuman viewership behavior, and next tries to prune out lockstep views from these broadcasts which make the broadcast appear more genuine. Our approach achieves over 98\% precision in identifying botted broadcasts and over 90\% precision/recall in identifying views in large viewbot attacks. 

\bibliographystyle{abbrv}
\bibliography{neil} 
\end{document}